\documentclass[prl,twocolumn,showpacs,preprintnumbers,amsmath,amssymb,superscriptaddress]{revtex4-1}

\usepackage{graphicx}
\usepackage{epsfig}
\usepackage{amsmath}
\usepackage{amssymb}
\usepackage{textcomp}

\usepackage{dcolumn}
\usepackage{bm}

\newcommand{\ket}[1]{|#1\rangle}

\begin{document}
\newcommand*{\MAINZ}{Institut f\"ur Quantenphysik,
Universit\"at Mainz, Staudingerweg 7, 55128 Mainz, Germany}
\newcommand*{\ULM}{Institut f\"ur Quanteninformationsverarbeitung,
Universit\"at Ulm, Albert-Einstein-Allee 11, 89081 Ulm, Germany}

\homepage{http://www.quantenbit.de}

\title{Observing the phase space trajectory of an entangled matter wave packet}
\author{U. Poschinger}\email{poschin@uni-mainz.de}\affiliation{\ULM}\affiliation{\MAINZ}
\author{A. Walther}\affiliation{\ULM}\affiliation{\MAINZ}
\author{K. Singer}\affiliation{\ULM}\affiliation{\MAINZ}
\author{F. Schmidt-Kaler}\affiliation{\ULM}\affiliation{\MAINZ}

\date{\today}

\begin{abstract}
We observe the phase space trajectory of an entangled wave packet of a trapped ion with high precision. The application of a spin dependent light force on a superposition of spin states allows for coherent splitting of the matter wave packet such that two distinct components in phase space emerge. We observe such motion with a precision of better than 9\% of the wave packet extension in both momentum and position, corresponding to a 0.8~nm position resolution. We accurately study the effect of the initial ion temperature on the quantum entanglement dynamics. Furthermore, we map out the phonon distributions throughout the action of the displacement force. Our investigation shows corrections to simplified models of the system evolution. The precise knowledge of these dynamics may improve quantum gates for ion crystals and lead to entangled matter wave states with large displacements.
\end{abstract}


\maketitle

Quantum entanglement of matter is a fascinating subject to study, as the most fundamental features of quantum physics become apparent, and as its observation in modern experimental realizations of former Gedanken experiments forces us to totally abandon any classical imagination of solid particles, or even of matter waves.

Beyond this fundamental interest, quantum entanglement is a useful resource for many important tasks, including information processing, communication, and precision measurements. In the last decade we have witnessed a rapid growth of experiments with well-defined atomic quantum systems, the most prominent among them being two-level atomic quantum systems denoted as qubits \cite{BLATT2008,RAIMOND2001,VION2002}. Quantum tomography is a diagnostic tool for investigating such quantum states, and has been used for systems of up to eight entangled ions \cite{HAEFFNER2005,RIEBE2006,HOME2009}.  For trapped ions, entangling quantum gates are based on the transient entanglement between internal qubit and external motional degrees of freedom, which is finally mapped back onto the qubit state, such that the motion is disentangled from the qubit after the operation. High fidelity gate operations are only possible if not the slightest trace of information is left in the motional degrees of freedom. Therefore, a high degree of control over the motion is required, for which in turn it is of fundamental interest to precisely monitor the dynamics of spin and motion.

The focus of this letter is the entanglement of the spin degree of freedom of a trapped ion with its motional quantum state under the dynamics of  laser driven displacement operations \cite{LEIBFRIED2005,HALJAN2005,MONROE1996,MCDONNELL2007}. We present an analysis of the laser-driven dynamics in both phase space and Hilbert space. We sense higher-order terms of the ion-light interaction Hamiltonian, and we are able to reveal experimentally why quantum superpositions of coherent states are increasingly difficult to prepare when the displacement magnitude becomes larger. As the demonstrated tomography scheme is applicable to determine the coherence of the entangled state \cite{DELEGLISE2008} - here for the fundamental system of a trapped ion crystal interacting with laser pulses for quantum gate operations - it is of fundamental importance for the  investigation of the scalability of quantum information processing based on trapped ions.

\begin{figure}[th!]
\begin{center}
\includegraphics[width=0.5\textwidth]{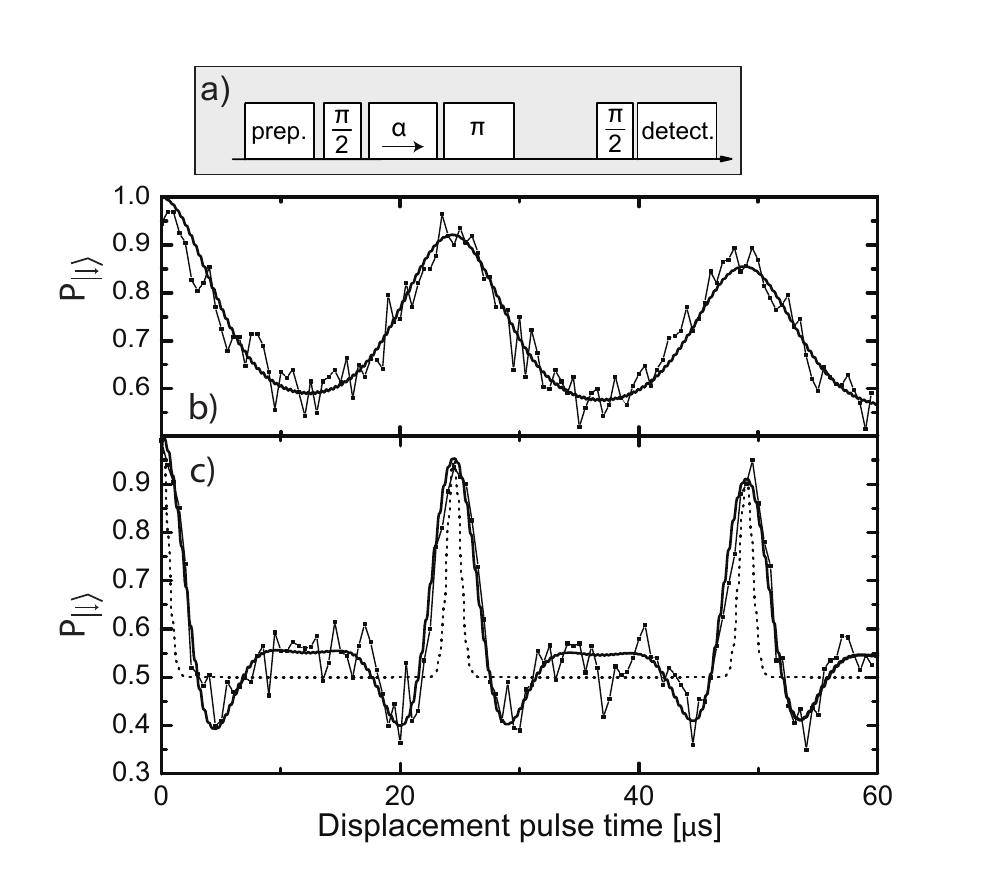}
\caption[]{Time evolution of entanglement and disentanglement for a ground state wave packet: (a) Experimental pulse sequence, see text. b) The phase contrast as measured versus duration of the displacement pulse taken for a ground state and (c) a Doppler cooled ion. The solid lines result from a fit to Eq.~\ref{eq:fringecontrast}. For the Doppler cooled ion, the fit parameters are identical to the ground state case except that a thermal mean phonon number $\bar{n}\approx$ 20 is assumed and thermal averaging was performed according to Eq. \ref{eq:fringecontrast2}. The dashed line indicates the prediction of Eq. 2 in Ref.\cite{HALJAN2005}, neglecting the nonclassical dependence of the force magnitude on the motional state.}
\label{fig:entangled_f1}
\end{center}
\end{figure}

This paper is organized as follows: after a brief introduction to our specific toolbox of ion quantum state preparation, we describe how to generate and manipulate entangled quantum states. We then present data which elucidates the role of the initial state preparation e.g. the ion temperature, for the dynamics caused by the spin-dependent force. We show complementary ways to analyze the dynamics of the optically driven ion; one being the measurement of the motional state by mapping out phonon distributions, i.e. we look at the dynamics in Hilbert space in Fock representation. A second approach maps out the dynamics in phase space, i.e. we follow the ion's trajectory with extremely high precision.

For the experiments we use a micro-structured Paul trap \cite{SCHULZ2006} which provides three dimensional harmonic confinement with frequencies of $\omega/(2\pi)$ = $\{$1.35, 2.4, 3$\}$ MHz for a single $^{40}$Ca$^+$ ion, where the lowest frequency $\omega_{ax}$ pertains to the axial vibrational mode. We apply Doppler cooling on the S$_{1/2}$ to P$_{1/2}$ transition near 397~nm. A magnetic field of about B = 0.4~mT splits both Zeeman qubit-levels of the ground state S$_{1/2}$, labeled $\{\ket{\uparrow}, \ket{\downarrow}\}$, by 18~MHz. We initialize the ion by sideband ground state cooling \cite{SCHULZ2008,POSCHINGER2009}, followed by optical pumping to the state $|\uparrow\rangle$. For coherent manipulations of the qubit we drive stimulated Raman transitions: the ion is irradiated with two laser beams near 397~nm at a detuning of $\Delta/(2\pi)$ = 40~GHz from the dipole transition. To perform different operations we utilize three different beam geometries: (i) Two co-propagating beams, R1 and CC, orthogonal to the direction of B with a relative detuning resonant to the Zeeman splitting, drive single qubit rotations without coupling to any motional degrees of freedom. (ii) Two beams, R1 and R2, with R2 $\bot$ R1 and R2 $\|$ B, where both are aligned at 45\textdegree with respect to the axial trap direction. Thus, the k-vectors of R1 and R2 establish a difference vector, $\delta k$, along the trap axis used for momentum transfer and excitation of axial vibration. Spin qubit rotations may be driven with a coupling to the ions axial mode, characterized by a Lamb-Dicke factor $\eta = \delta k \hspace{1mm} x_0\approx$ 0.25, where $x_0$ is the ground state wave packet extension. This geometry is used for Raman sideband cooling and to drive Rabi oscillations on motional sidebands. (iii) A third pair of beams is comprised of R2 and CC, where only circular light polarization components are present such that no coupling of the spin qubit levels occurs, but axial ion motion can be excited via an ac-Stark light force oscillating close to the vibrational frequency. Phase and magnitude of this drive depend on the spin state of the ion \cite{MONROE1996,HALJAN2005}. After performing manipulations on the spin and the motional state, the spin is read out by transferring the population from $\ket{\uparrow}$ to the metastable D$_{5/2}$ state via a rapid adiabatic passage pulse \cite{WUNDERLICH2007,POSCHINGER2009}. When illuminated with resonant light near 397~nm, the ion is measured to be in $\ket{\downarrow}$ if we detect fluorescence, and it is measured to be in D$_{5/2}$, corresponding to $\ket{\uparrow}$, if it remains dark. We repeat the sequence 200 times to determine the spin occupation probabilities $P_{\uparrow}$ and $P_{\downarrow}$.

\begin{figure*}[]
\begin{center}
\includegraphics[width=0.9 \textwidth]{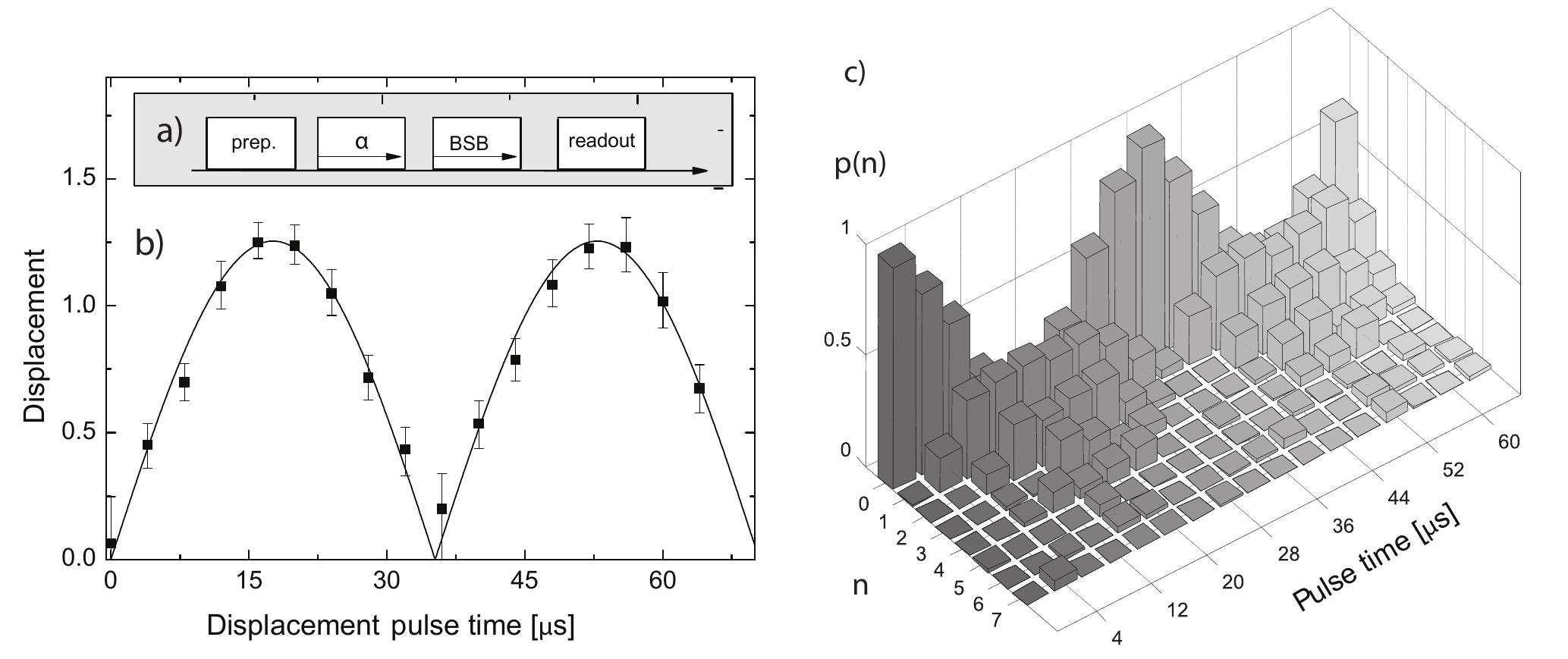}
\caption[]{(Phonon distribution dynamics: \textbf{a)} The experimental measurement sequence, see text. \textbf{b)} Measured displacement parameter versus displacement pulse time obtained from the measured phonon distributions, the solid line is a fit to Eq.~\ref{eq:alphat}. \textbf{c)} Reconstructed phonon distributions from the blue sideband Rabi oscillations. The typical confidence intervals for the are up to $\pm$0.1 for the displaced states and about -0.05 for the states with small or no displacement.}
\label{fig:entangled_f2}
\end{center}
\end{figure*}

We generate entangled wave packets with the sequence in Fig.\ref{fig:entangled_f1}(a), where the spin echo sequence of $\pi/2$-pulse, $\pi$-pulse, and $\pi/2$-pulse is formed with Raman interactions of type (i) and the spin-dependent displacement employs the Raman interaction of type (iii). With a relative detuning of the R2 and CC beams of $\delta_{R2,CC}=\omega_{ax}-\delta$, acting for time $t$ on a superposition state, the resulting state is
\begin{equation}
|\Psi_f\rangle=\frac{1}{\sqrt{2}}\left(|\uparrow,\alpha(t)\rangle+i|\downarrow,-\alpha(t)\rangle\right),
\end{equation}
with the displacement \cite{LEIBFRIED2003}
\begin{equation}
\alpha(t)= -\frac{\eta\Delta_S}{2\delta}e^{i\delta t/2}\sin\frac{\delta t}{2},
\label{eq:alphat}
\end{equation}
where $\Delta_S$ is a differential ac-Stark shift arising from the laser beams R2 and CC. When the concluding $\pi/2$-pulse acts on this state, the spin will only flip completely into $\ket{\uparrow}$ if no displacement was present, since this can be seen as a which-path information, suppressing the spin interference signal. The fringe contrast is given by the overlap of the adjacently displaced ground state wave packets:
\begin{equation}
\mathcal{C}(t)=|\langle -\alpha(t)|\alpha(t)\rangle|^2 = e^{-2|\alpha(t)|^2}.
\label{eq:fringecontrast}
\end{equation}
Thus, the observed contrast serves as a measure of the displacement magnitude. As the motion is driven slightly off-resonant with $\delta_{R2,CC}$, both spin components become periodically entangled and disentangled, and we observe the coherence decay and revive, correspondingly. The data plotted in Fig.~\ref{fig:entangled_f1} (a) show the dynamics of $P_{\uparrow}(t)$ at the end of the sequence, indicating that the wave packets of both spin components are driven back into the origin after about 24~$\mu$s, and again near 48~$\mu$s, as expected from $\delta = 2\pi\cdot42$~kHz.

For applications of the displacement operation, it is of interest to investigate the effect of imperfect preparation of the motional state, i.e. an initial thermal excitation. Fig.~\ref{fig:entangled_f1} (b) shows the contrast signal for a Doppler cooled ion with an average phonon number $\bar{n}\approx 20$. In order to accurately describe all features of this data, we develop a model that includes the fact that the displacement force also depends non-classically on the motional state. As a result, the expression for the contrast that we obtain by a thermal averaging of Eq.~\ref{eq:fringecontrast}, includes displacements $\alpha_n(t)$ that depend explicitly on $n$:
\begin{equation}
\mathcal{C}(t)=\sum_n p_n^{(i)} |\langle n,-\alpha_n(t)|n,\alpha_n(t)\rangle|^2.
\label{eq:fringecontrast2}
\end{equation}
Here, $p_n^{(i)}$ is the initial thermal phonon distribution, and the state evolutions, $\alpha_n(t)$, are determined by a quantum dynamical simulation. The resulting average of Eq.~\ref{eq:fringecontrast2} is plotted as the solid line in Fig.~\ref{fig:entangled_f1}~(c), and is in very good agreement with the experimental data. Previously used models, such as the one in Ref.~\cite{HALJAN2005}, do not include the $n$-dependence of the displacement force, and we see from the dashed line Fig.~\ref{fig:entangled_f1} (b) that such a model does not reproduce all features of the data. In particular, the side peaks next to the revival peaks can only be reproduced by including the $n$-dependence. The consideration of the dependence of the force magnitude on $n$ is of importance for multi-ion entangling gate operations in the thermal regime \cite{KIRCHMAIR2009a,KIRCHMAIR2009b}, where the thermal dispersion of trajectories can be among the main sources of infidelity.

The light-force mediated entanglement operation can be investigated not only in terms of spin coherences, but also by directly monitoring the motional degree of freedom. We drive the displacement dynamics and examine the resulting motional state with the sequence in the inset of Fig.~\ref{fig:entangled_f2} by using the Raman beams in configuration type (ii), where resonant Rabi oscillations on the blue motional sideband (bsb) contain information on the motional state, i.e. the occupation probabilities, $p_n$, of the Fock states $|n\rangle$ \cite{MEEKHOF1996,BRUNE1996}. The anti-Jaynes-Cummings like resonant bsb excitation for a time $t_p$ results in a signal
\begin{equation}
P_{|\downarrow\rangle}(t_p) =\sum_n\frac{p_n}{2}\left(a\;e^{-t/\tau}\cos\left(\Omega_{n,n+1}\;t_p\right)+1\right),
\label{eq:bsbsignal}
\end{equation}
from which we may obtain the $p_n$ as the spectral components at Rabi frequencies $\Omega_{n,n+1}\approx\eta\sqrt{n+1}\;\Omega_0$. As a more precise way to obtain the $p_n$ along with the parameters $\Omega_0,a$ and the decoherence time $\tau$, we employ a maximum-likelihood reconstruction by means of a genetic algorithm. The resulting phonon distributions for the various displacement pulse times can be fit to distributions pertaining to a coherent state, $p_n(\alpha) = e^{-|\alpha|^2} |\alpha|^{2n}/n!$. The resulting values  $\alpha(t)$ are shown in Fig.~\ref{fig:entangled_f2} (a), where one can clearly observe the periodical excursion of the wave packet. The measured phonon distributions in Fig.~\ref{fig:entangled_f2} (b) indicate even more strikingly how the light force driven motion returns the ion back to the vibrational ground state near times of 30 $\mu$s.

\begin{figure}[t]
\includegraphics[width=0.5\textwidth]{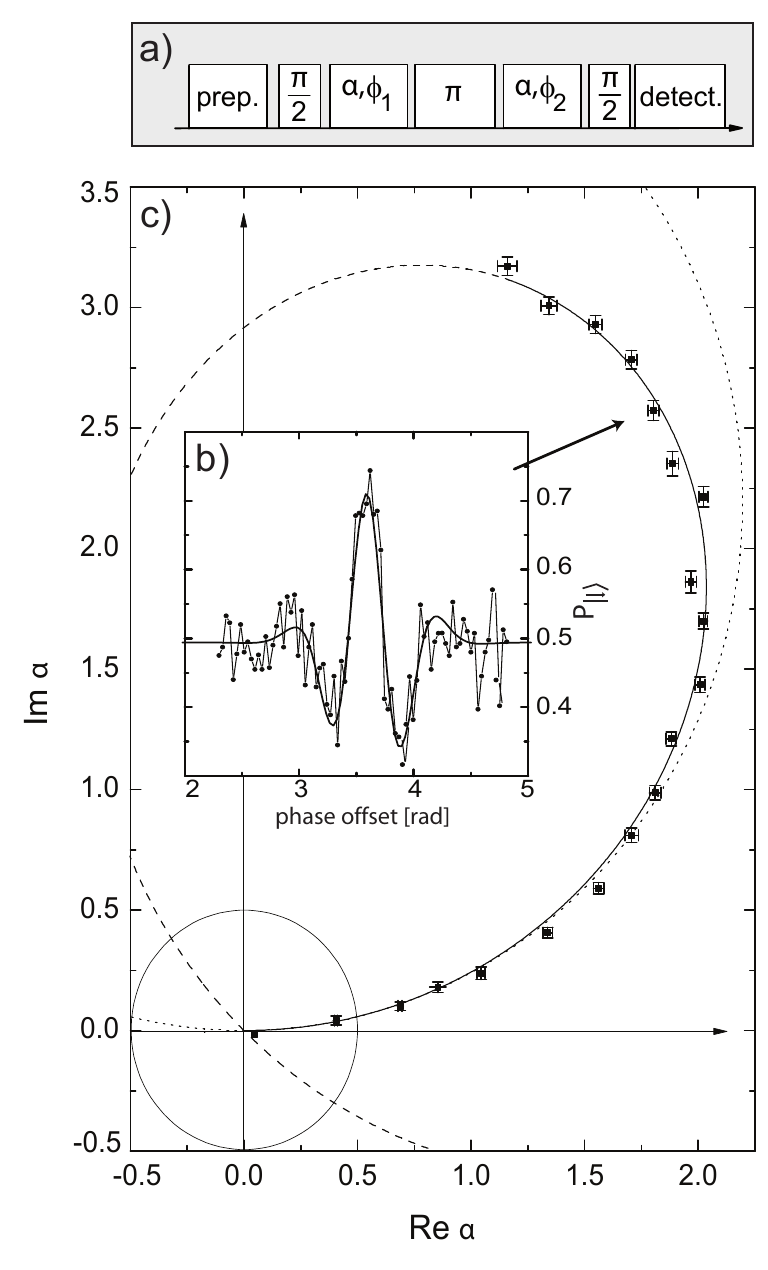}
\caption[]{(a) Trajectory measurement pulse scheme, see text. (b) Homodyne wave function beat signal, for a displacement pulse of length t=60$\mu$s along with a fit to Eq.~\ref{eq:beatsignal}. (c) The resulting phase space coordinates $|\alpha|e^{i\phi}$ inferred from fitting the measured signals $P_{\uparrow}(\phi)$ to Eq.~\ref{eq:beatsignal}, along with the theoretical trajectory Eq.~\ref{eq:alphat}. The outer dashed circle indicates the trajectory that would be observed in the case of a spatially homogeneous force. The circle around the origin indicates the $e^{-1/2}$ radius of the Wigner function, more than ten times larger than our maximum measurement errors in $\alpha$.}
\label{fig:entangled_f3}
\end{figure}

For an ultra-precise determination of the wave packet dynamics, we employ a wave packet homodyning technique to map out the dynamics in phase space \cite{MONROE1996}, see Fig.~\ref{fig:entangled_f3}(a). The spin superposition state formed by the $\pi/2$-pulse is affected by the light force from the type (ii) Raman interaction such that only the $\ket{\downarrow}$ component is shifted by $\alpha e^{-i\delta t/2 + \phi_1}$ while $\ket{\uparrow}$ remains at the origin of phase space. The Raman type (i) driven $\pi$-pulse flips both spin states, and now the displacement Raman type (ii) light force acts on that wave packet component which was left before at $\alpha$=0. We have chosen the amplitudes to be equal, but the phase $\phi_2$ of the second  drive is varied such that both spin components only partially overlap, depending on the difference $\Delta \phi = \phi_2 - \phi_1$. When the spin echo sequence is concluded by the last $\pi/2$-pulse, the width and the phase of the interference pattern, Fig.~\ref{fig:entangled_f3}(b), allow for determining the magnitude and phase of $\alpha$. The upper spin state occupation probability finally reads \cite{MONROE1996}:

\begin{equation}
P_{\uparrow}(t)=\tfrac{1}{2}\left(1-e^{-|\alpha|^2(1-\cos\phi(t))-t/\tau}\cos\left(|\alpha|^2\sin\phi(t)\right)\right).
\label{eq:beatsignal}
\end{equation}

Here $\phi(t)=\Delta\phi+\delta t+\delta t_w$ is the harmonic oscillator phase picked up during the driving($t$) and idle($t_w$) times in the sequence.  We also empirically include laser-induced decoherence by an exponential decay factor $\exp(-t/\tau)$, which is predominantly caused by the fluctuating ac-Stark shift from intensity noise of the circularly polarized R2 beam. Sets of $P_{\uparrow}$ were recorded for displacement pulse durations ranging from 0~$\mu$s to 76~$\mu$s in steps of 4~$\mu$s for varying beat phases $\Delta\phi$, and the measurement data is fit to Eq.~\ref{eq:beatsignal}.

The resulting real and imaginary parts of the displacement $\alpha(t)$  are plotted in  Fig.~\ref{fig:entangled_f3}(c). We can clearly identify deviations from the idealized dynamics, as the wave packet excursion approaches the wavelength of the driving light wave and the Lamb-Dicke approximation fails. The phase $\phi(t)$ is given by the center of the envelope in the beat signal. From the time dependency of $\phi_0(t)$, we reveal the detuning, $\delta$, with high precision, and it is found to be $2\pi\cdot$5.237(27)~kHz. Thus, we determine the vibrational frequency with a relative accuracy on the order of 10$^{-5}$. Our homodyne measurement scheme can be seen as a continuous variable analog of Ramsey spectroscopy. We reach a very high accuracy, and a comparable performance with conventional Ramsey pulses would require long delay times of tenths of ms.

From the data plotted in Fig.~\ref{fig:entangled_f3}(c), one can clearly recognize the deviation from the circular trajectory predicted by Eq.~\ref{eq:alphat}. This can be accounted for by empirically introducing an effective return time $t_{\text{ret}}^{\text{eff}}$ \cite{MCDONNELL2007}, corresponding to an effective detuning $\delta_{\text{eff}}\not=\delta$ in the argument of the sine function in Eq.~\ref{eq:alphat}. A fit to the modified Eq.~\ref{eq:alphat} reveals an effective detuning of $\delta_{\text{eff}}=2\pi\cdot$6.63(10)~kHz. The trajectory can be reconstructed from the values for $|\alpha(t)|$ and $\phi(t)$, as it is shown in Fig.~\ref{fig:entangled_f3}, up to an unknown angle of rotation around the origin. This angle is given by the relative optical phase between the R2 and CC beams at the ion location, which is varying from shot to shot. The measurement accuracies along both axes are much smaller than the dimension of the ground state wave packet size. This does of course not violate the Heisenberg uncertainty principle, as the measurement is statistical and its accuracy relies on the shot-to-shot reproducibility. The quantum simulation e.g. of quantum random walks \cite{SCHMITZ2009,KARSKI2009,ZAEHRINGER2010} or the Dirac equation \cite{GERRITSMA2010} may benefit from applying our method for a precise observation of the wave packet dynamics.

In conclusion, we were able to precisely follow the phase space trajectory of the entangled wave packets and study in detail decoherence and dephasing effects of such states. Our experimental investigation of quantum interferences illustrates the sources which naturally and quite in general limit the observation of quantum entanglement. We could show that a significant contribution to the wave packet dynamics comes from higher order (non-homogeneous) terms of the interaction Hamiltonian. This leads to more complicated trajectories in phase space, as seen by the deviation from the predicted trajectory, and this behavior becomes increasingly important the larger the displacement is. In the future, we plan to overcome such limits by applying a temporally tailored light force which takes into account the system evolution for creating highly non-classical entangled matter waves, by modulating the phase and amplitude of the drive fields. Furthermore, in our experiments we could show how the initial temperature of the system affects the light-force drive and describe this by the correct theoretical model. This could be of crucial importance for devising gate schemes which are more robust than the current ones.

We acknowledge financial support by the European commission within EMALI and the STREP-MICROTRAP, the IPs SCALA and AQUTE.

\vspace{1cm}

\end{document}